\documentclass[sn-mathphys-num]{sn-jnl}% Math and Physical Sciences Numbered Reference Style 
\usepackage{graphicx}%
\usepackage{multirow}%
\usepackage{amsmath,amssymb,amsfonts}%
\usepackage{amsthm}%
\usepackage{mathrsfs}%
\usepackage[title]{appendix}%
\usepackage{xcolor}%
\usepackage{textcomp}%
\usepackage{manyfoot}%
\usepackage{booktabs}%
\usepackage{algorithm}%
\usepackage{algorithmicx}%
\usepackage{algpseudocode}%
\usepackage{listings}%
\usepackage{pifont} %勾叉
\usepackage{subfigure} %子图
\usepackage{lineno}  %行号
\usepackage{CJK} %中文 \begin{CJK*}{UTF8}{gkai}李志军\end{CJK*}

\theoremstyle{thmstyleone}%
\theoremstyle{thmstyletwo}%

\theoremstyle{thmstylethree}%

\begin{document}
% \setrunninglinenumbers
% \begin{linenumbers}

\title[Article Title]{\boldmath Visualization for physics analysis improvement and applications in BESIII}

%\iffalse

 \author[1]{\fnm{Zhi-Jun} \sur{Li} (\begin{CJK*}{UTF8}{gkai}李志军\end{CJK*})}
 \author[2,1]{\fnm{Ming-Kuan} \sur{Yuan} (\begin{CJK*}{UTF8}{gkai}袁铭宽\end{CJK*})}
 \author[3,4]{\fnm{Yun-Xuan} \sur{Song} (\begin{CJK*}{UTF8}{gkai}宋昀轩\end{CJK*})}
 \author[4]{\fnm{Yan-Gu} \sur{Li} (\begin{CJK*}{UTF8}{gkai}李彦谷\end{CJK*})}
 \author[1]{\fnm{Jing-Shu} \sur{Li} (\begin{CJK*}{UTF8}{gkai}李静舒\end{CJK*})}
 \author[5,6]{\fnm{Sheng-Sen} \sur{Sun} (\begin{CJK*}{UTF8}{gkai}孙胜森\end{CJK*})}
 \author[2]{\fnm{Xiao-Long} \sur{Wang} (\begin{CJK*}{UTF8}{gkai}王小龙\end{CJK*})}
 \author*[1]{\fnm{Zheng-Yun} \sur{You} (\begin{CJK*}{UTF8}{gkai}尤郑昀\end{CJK*})}\email{youzhy5@mail.sysu.edu.cn}
 \author[4]{\fnm{Ya-Jun} \sur{Mao} (\begin{CJK*}{UTF8}{gkai}冒亚军\end{CJK*})}

 \affil[1]{\orgdiv{School of Physics}, \orgname{Sun Yat-sen University}, \orgaddress{\city{Guangzhou}, \postcode{510275}, \country{China}}}

 \affil[2]{\orgdiv{Institute of Modern Physics}, \orgname{Fudan University}, \orgaddress{\city{Shanghai}, \postcode{200433}, \country{China}}}

 \affil[3]{\orgdiv{Institute of Physics}, \orgname{\'Ecole Polytechnique  F{\'e}d{\'e}rale de Lausanne (EPFL)}, \orgaddress{\city{Lausanne}, \postcode{1015}, \country{Switzerland}}}

 \affil[4]{\orgdiv{School of Physics}, \orgname{Peking University}, \orgaddress{\city{Beijing}, \postcode{100871}, \country{China}}}

 \affil[5]{\orgdiv{Institute of High Energy Physicss}, \orgname{Chinese Academy of Sciences}, \orgaddress{\city{Beijing}, \postcode{100049}, \country{China}}}

 \affil[6]{\orgname{University of Chinese Academy of Sciences}, \orgaddress{\city{Beijing}, \postcode{100049}, \country{China}}}

 % \affil[7]{\orgdiv{Department}, \orgname{Organization}, \orgaddress{\street{Street}, \city{City}, \postcode{10587}, \state{State}, \country{Country}}}


 %\affil[2]{\orgdiv{Department}, \orgname{Organization}, \orgaddress{\street{Street}, \city{City}, \postcode{10587}, \state{State}, \country{Country}}}
 
%\fi


\abstract{Modern particle physics experiments usually rely on highly complex and large-scale spectrometer devices. 
In high energy physics experiments, visualization helps detector design, data quality monitoring, offline data processing, and has great potential for improving physics analysis.
In addition to the traditional physics data analysis based on statistical methods, visualization provides unique intuitive advantages in searching for rare signal events and reducing background noises. 
By applying the event display tool to several physics analyses in the BESIII experiment, we demonstrate that visualization can benefit potential physics discovery and improve the signal significance.
With the development of modern visualization techniques, it is expected to play a more important role in future data processing and physics analysis of particle physics experiments.}




\maketitle

\section{Introduction}
\label{sec:introduction}
Visualization has always been an essential technique since the birth of nuclear and particle physics.
In the early days of the 20th century, when the field of particle physics emerged, nuclear emulsion, cloud chambers, and bubble chambers were used to detect particles. The specific particles and event processes are observed directly with a high level of intuitiveness during the data analysis. With the advancement of particle physics experimental techniques, this visual observation method can no longer meet the increasing demands for higher precision and larger data volume for modern particle physics experiments. Consequently, high-speed electronic data processing and
statistical-based analysis methods become popular for massive data processing.

Nowadays, large-scale general-purpose spectrometers such as ATLAS~\cite{ATLAS:2008xda}, CMS~\cite{CMS:2008xjf}, LHCb~\cite{LHCb:2008vvz,LHCb:2014set}, BESIII~\cite{Ablikim:2009aa}, BaBar~\cite{BaBar:2001yhh}, Belle~\cite{Belle:2000cnh} and BelleII~\cite{Belle-II:2010dht}, as well as other collider based experiments, are playing significant roles in the field of particle physics.
High-precision large-scale spectrometers exhibit considerable complexity and often translate particle information into a set of non-intuitive electronic signals. 
%Researchers frequently need to analyze these electronic signals to understand the physical characteristics of the events. 
Visualization of large-scale spectrometers can recreate the behavior of particles in the detector more intuitively, allowing the spectrometers to maintain high speed and high-precision data processing while still being visually accessible~\cite{Bellis:2018hej,HEPSoftwareFoundation:2017ggl}.
However, visualization in particle physics applications often remains at the level of visual display, and the potential value of its application in specific physics analysis is usually overlooked.

In this study, we present the visualization-based method to assist physics analysis. 
Building on traditional analysis methods, the visualization-based analysis method can further find new discrimination variables, suppress the background, and quickly assess the quality of observed data in modern particle physics experiments.
BESIII, an electron-positron collider experiment with low background level, is suitable to demonstrate the advantages of the visualization method for physics analysis.
Taking BESIII as an example, the BESIII Visualization software (BesVis) will be introduced and its application in several specific physics analyses will be discussed.
%including the search for the invisible decay of $\Lambda$ baryon, the search for the semi-muonic charmonium decay $J/\psi\to D^{-}\mu^{+}\nu_{\mu}+c.c.$, the search for the lepton flavor violating decay $\psi(2S)\to e^{-}\mu^{+}+c.c.$ and the study of the semi-leptonic decay $\bar{\Lambda}^-_c\to \bar{n} l^{-}\bar{\nu}_{l}+c.c.$, where $c.c.$ means charge conjugation.

The remainder of this paper is organized as follows.
In Section~\ref{sec:Methodologies}, the analysis method based on visualization is presented.
In Section~\ref{sec:bes3}, we introduce the BESIII experiment and its visualization software.
The application of visualization to assist BESIII analyses is introduced in Section~\ref{sec:application}.
Finally, the potential for further development and applications is discussed in Section~\ref{sec:future}.

\section{Methodologies}
\label{sec:Methodologies}

Modern particle physics data analyses usually rely on statistical methods based on selection criteria to screen physically interesting signal events. The raw hits in the detector from real experiments or simulations are the original information, which will be reconstructed as an event or multi-events based on the detector geometry. The position and time of hits, cluster of multiple hits, particle vertex, momentum and energy of track, which are considered as low-level input information, can be obtained from the raw hits with the reconstruction to construct the physics event data. Based on the reconstruction data, more physics variables are necessary for further analysis, such as the invariant mass of particles, missing transverse energy (MET), missing momentum ($P_{\rm{miss}}$), kinematic fit updated four-momentum~\cite{Yan:2010zze}, which are treated as high-level event information. Events from different processes have different features on these physics variables, thus providing a portal to distinguish the signal and background in an analysis, which is usually called ``cut" in particle physics analysis. The cut from these statistical data distributions is referred to as the statistical cut-based method in this paper. With the statistical cut-based method, analyzers can apply cuts to reduce background and extract signals, processing a massive number of events and quantifying the statistical features of these events with great success in modern particle physics research. 
However, this method typically requires a certain number of event statistics to reveal the characteristics of the physics variables and usually depends on empirical choices for constructing suitable physics variables, which generally lack an intuitive understanding of the global features of events in the detector.
Before applying the statistical cut-based event selection, construction of the high-level physics variables also involves triggering, simulation, and reconstruction, in which the low-level event information, such as raw hits, tracks, and their association with the detector geometry, may have been lost in the stages of data processing.   

The visualization method has been widely used in the early development of particle physics, such as the observation of the strange particle by W. B. Fowler~\cite{Fowler:1954nq} and the first observation of ternary and quaternary fission of Uranium nuclei by Sanqiang Qian and Zehui He~\cite{sanqiang1947}.
Different from the multi-event, empirical, and software-dependent nature of the statistical cut-based method, the visualization method allows researchers to grasp the entire response of an event in the detector as a whole picture. It offers strong visual interpretability, with less reliance on empirical knowledge and data processing software, and enables a rapid examination of the physical characteristics of individual events. In modern particle physics, the statistical cut-based method is the basic data analysis method, while the visualization method can help further improve the physics analysis by overcoming the limitations of only using high-level event information with the statistical cut-based method. The visualization method is a beneficial approach to complement the statistical cut-based method, as shown in Table~\ref{tab:vs}.
%, which are not opposite to each other. 

The visualization method relies on three key inputs from the statistical cut-based method: low-level input from raw hits and detector, input from reconstruction (RECO) data, and events type input from the signal-background distinction. 
Based on these inputs, there are four important outputs for the statistical cut-based method that can help improve the physics analysis:
\begin{itemize}
\item Find reconstruction problem. By displaying the raw hits, detector and reconstruction information at the same time, it is easy to check whether the reconstruction is consistent with the raw hits, which is widely applied in data quality monitoring system and data reconstruction. Sec. \ref{sec:invisible} is an example of its application in physics analysis.
\item Eliminate fake signal. The remaining signals after the statistical cut-based method also need to be further checked with the visualization method event by event, especially in discovery of a rare process or new physics signals. Any performance that exceeds expectations in the visualization may indicate that there are no real signals but unexpected backgrounds, which can be further eliminated with the new statistical cuts in addition to the existing ones. Sec. \ref{sec:invisible} is an example of eliminating fake signals based on the visualization method with new cuts.
\item Understand background. The remaining background after statistical cut-based method can be further checked with the visualization method at the event level, which can illustrate the origin and behavior of the background, leading to the discovery of more approaches to further suppress the background. Sec. \ref{sec:Dmunv} is an example of studying the remaining background based on the visualization method.
\item Discover more useful cuts. The ``cut" variables in the statistical cut-based method are usually based on the experience, but the event display based on the visualization method can show the overall information of an event, helping reveal additional useful cut variables or new input for the cut variables, particularly those related to the detector structure or event reconstruction. In Sec. \ref{sec:emu} and Sec. \ref{sec:nenv}, we introduce the examples of discovering new cuts based on the visualization method.
\end{itemize}

\begin{figure*}[tbp]
\centering
%\subfigure[]
{\includegraphics[width=0.9\textwidth]{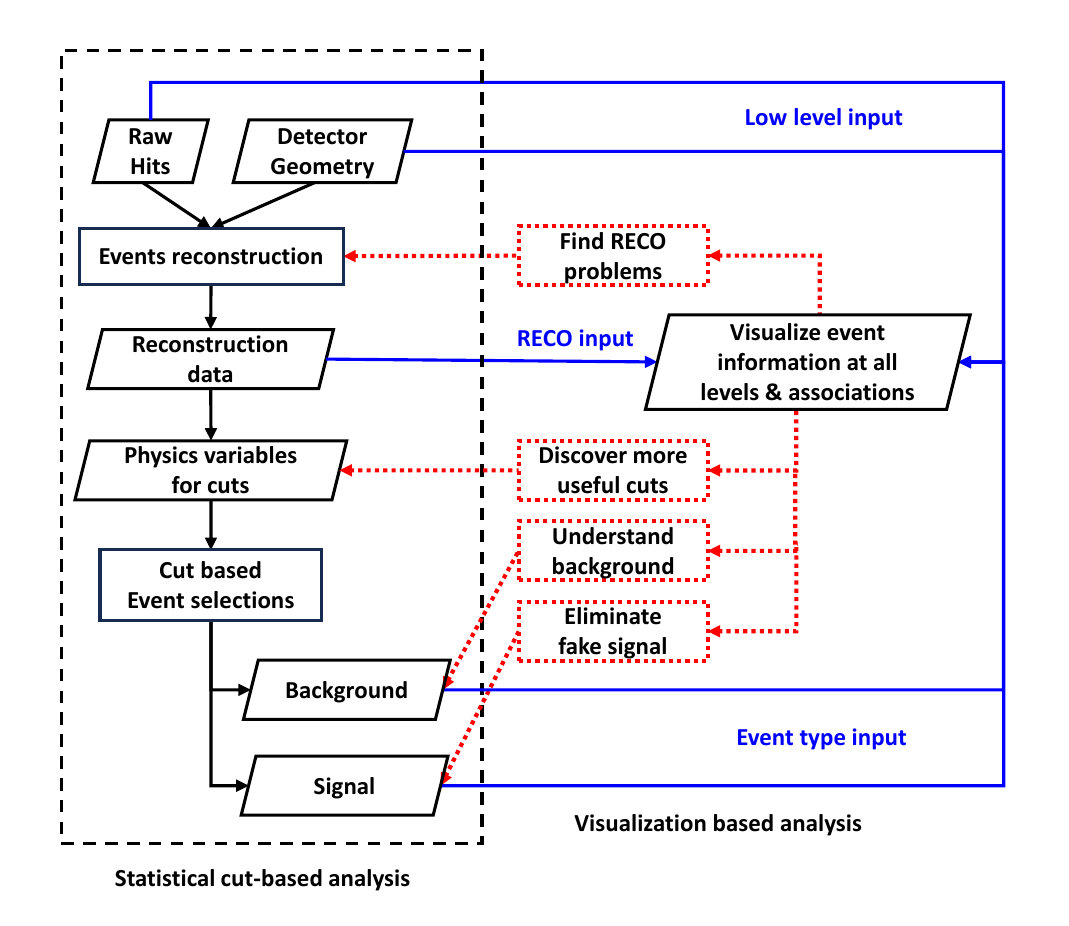}}\\
\hspace{0.03\textwidth}
\caption{The steps in general statistical cut-based physics analysis (left part) and the supplement of visualization in each step (right part). The two methods can be combined to improve the physics analysis.}
\label{fig:methodologies}
\end{figure*}

Figure \ref{fig:methodologies} shows the steps in general statistical cut-based physics analysis and the supplement of
visualization in each step, where the development of cuts can be further derived from the event characteristics obtained through visualization.
Up to now, the visualization method has been effectively applied in data quality monitoring systems, physics simulations, and data reconstruction. Additionally, due to the intuitiveness of event display software, visualized images are increasingly used in public news, journal covers, and outreach articles to present the physics processes visually. However, direct application of visualization in specific physics analysis is still limited.

%%%%%%%%%%%%%%%%%%%
%%%%%%%%%%%%%%%%%%%

%%%%%%%%%%%%%%%%%%%%%%%%%%%%%%%%%%%%
\begin{table*}[!htbp]
\caption{The advantages and disadvantages of statistical cut-based and visualization analysis.}
\setlength{\abovecaptionskip}{1.2cm}
\setlength{\belowcaptionskip}{0.2cm}
\begin{center} 
\footnotesize
\vspace{-0.0cm}
\begin{tabular}{l|cc}
\hline \hline
		%Characteristic & Statistical cut-based analysis & Visualization analysis\\
            Characteristic & Statistical cut-based analysis & Visualization\\
		\hline
		Processing a large number of events & \ding{51} & \ding{55} \\
            Quantifying the statistical features of multiple events & \ding{51} & \ding{55} \\
            Relying on other software and experience & \ding{51} & \ding{55} \\
            Highly intuitive & \ding{55} & \ding{51} \\
            %Obtaining comprehensive detailed information for a single event. & \ding{55} & \ding{51} \\
            Comprehensive detailed information for a single event & \ding{55} & \ding{51} \\
\hline \hline
\end{tabular}
\label{tab:vs}
\end{center}
\end{table*}
\vspace{-0.0cm}
%%%%%%%%%%%%%%%%%%%%%%%%%%%%%%%%%%%%

\section{BESIII and visualization}
\label{sec:bes3}

%%%%%%%%%%%%%%%%%%%
\begin{figure*}[tbp]
\centering
%\subfigure[]
{\includegraphics[width=0.9\textwidth]{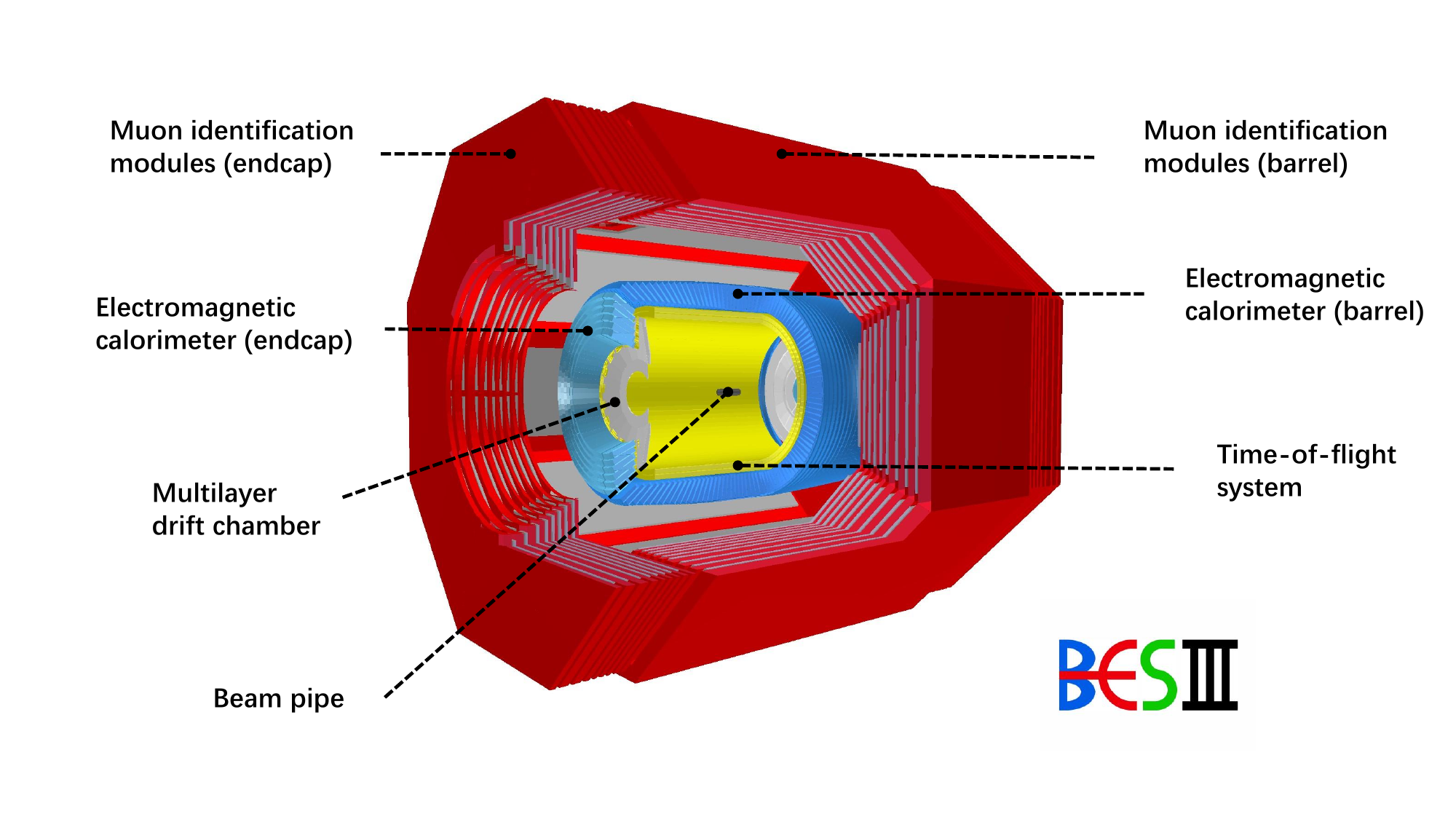}}\\
\hspace{0.03\textwidth}
\caption{Visualization of BESIII detector, where the outermost red part interleaved with white layers is MUC, the blue part is EMC, the yellow part is TOF, the light grey part is MDC, and the central dark grey part is the beam pipe.}
\label{fig:detector}
\end{figure*}
%%%%%%%%%%%%%%%%%%%

Beijing Spectrometer III (BESIII)~\cite{Ablikim:2009aa} is a general-purpose spectrometer for $\tau$-charm physics study in the center-of-mass energy range from 2.0 to 4.7~GeV. BESIII records symmetric $e^+e^-$ collisions provided by the Beijing Electron Positron Collider II (BEPCII) storage ring~\cite{Yu:IPAC2016-TUYA01} and has collected large data samples in this energy region~\cite{BESIII:2020nme}. The cylindrical core of the BESIII detector covers 93\% of the full solid angle and consists of a helium-based multilayer drift chamber~(MDC), a plastic scintillator time-of-flight system~(TOF), and a CsI(Tl) electromagnetic calorimeter~(EMC), which are all enclosed in a superconducting solenoidal magnet providing a 1.0~T magnetic field. The solenoid is supported by an octagonal flux-return yoke with resistive plate counter muon identification modules interleaved with steel~(MUC). 
Visualization of the BESIII detector is shown in Figure~\ref{fig:detector}.

\begin{figure*}[tbp]
\centering
\subfigure[]
{\includegraphics[width=0.9\textwidth]{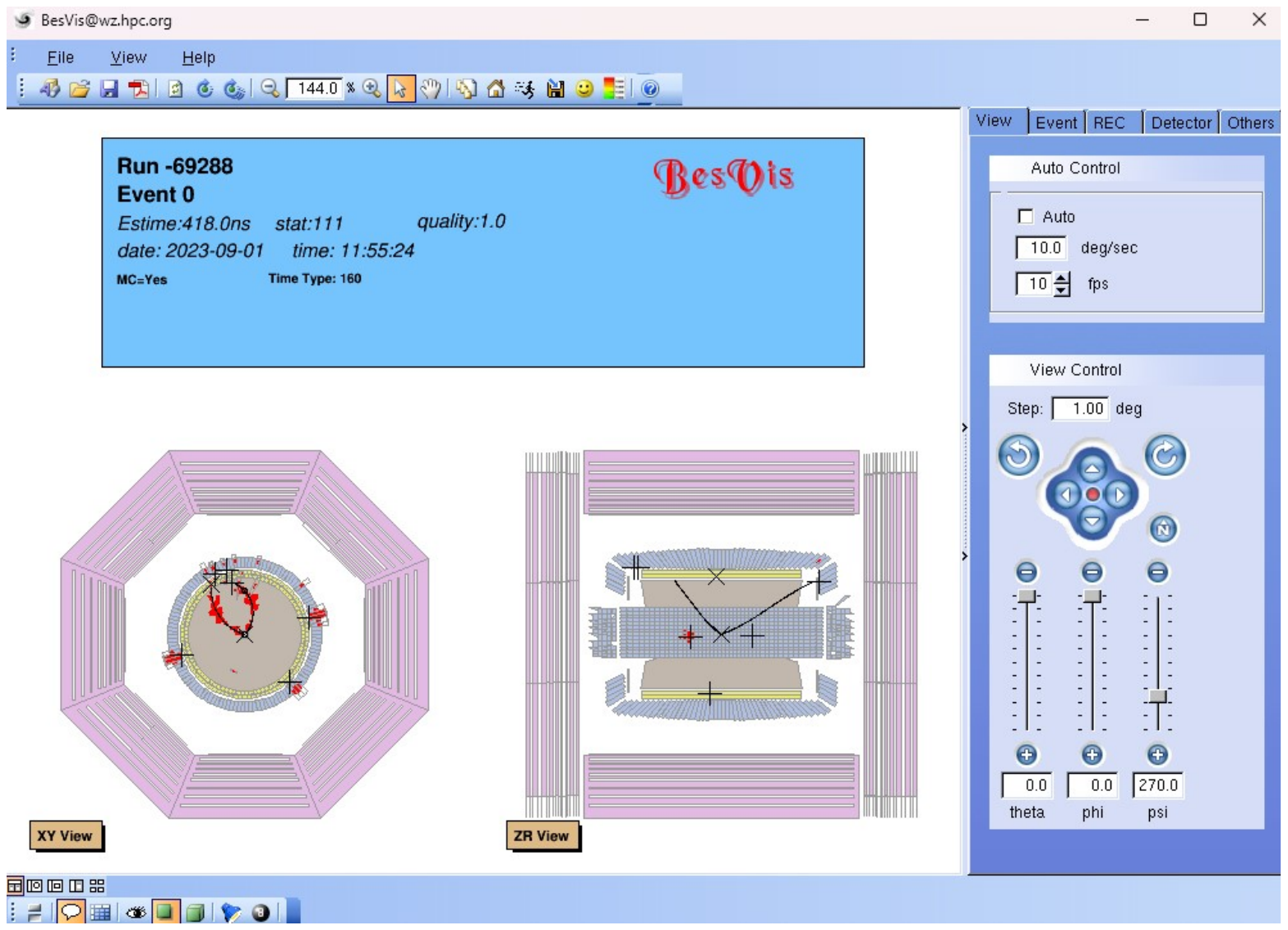}}\\
\hspace{0.0\textwidth}
\subfigure[]
{%\includegraphics[width=0.45\textwidth]{figure/etac2GG_3D_1.png}
 \includegraphics[width=1.0\textwidth]{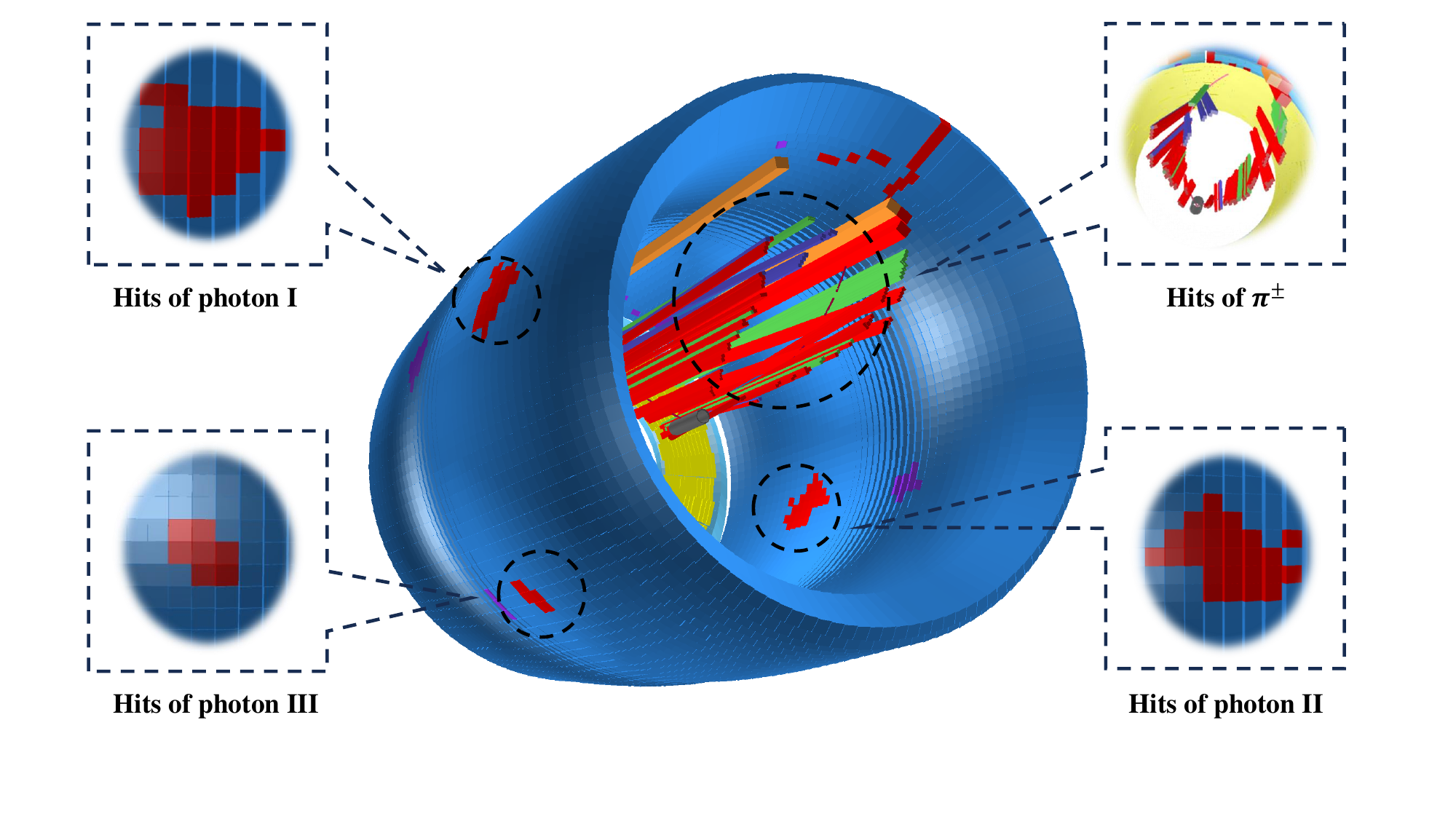}
}\\
\caption{GUI of BesVis and display of an example event with 2D (a) and 3D (b) visualization, where the event is $\psi(2S)\to\pi^{+}\pi^{-}J/\psi, J/\psi\to\gamma\eta_c, \eta_c\to\gamma\gamma$ from simulation. 
%(a) is the two-dimensional display. From outer to inner, the pink ring represents the MUC, the blue ring represents the EMC, the yellow ring represents the TOF, and the innermost gray area represents the MDC. The red color represents the hits of particles in the detector, and the black tracks and crosses represent reconstructed particle information. The two circular tracks in the MDC represent $\pi^{+}\pi^{-}$ and the EMC shows three red hits that are not pipi interactions, corresponding to three photons.
%(b) is the three-dimensional display. To emphasize the visualization of the event, some sub-detectors are turned off. The hits on the two barrel tracks indicate $\pi^{+}\pi^{-}$ interactions. The blue EMC barrel wall shows three red hits that are not $\pi^{+}\pi^{-}$ interactions, corresponding to three photons, where the two large clusters correspond to high-energy decay photons of $\eta_c$, and the small cluster corresponds to low-energy radiated photons of $J/\psi$.
}
\label{fig:2D 3D}
\end{figure*}

BesVis~\cite{LIANG2009325,Huang_2018} is a visualization software in the BESIII experiment, which is developed with ROOT~\cite{BRUN199781} in the BESIII Offline Software (BOSS). As shown in Figure~\ref{fig:2D 3D} (a), its graphical user interface (GUI) is based on ROOT GUI package and the geometric description is provided in the format of Geometry Description Markup Language (GDML)~\cite{LIANG2009325}. By reading GDML files, BesVis generates detector geometry in ROOT format, ensuring complete consistency with the detector description used in the simulation and reconstruction of offline software. 
%This ensures the accuracy and consistency of the geometry throughout the entire simulation and reconstruction. 

BesVis supports both two-dimensional and three-dimensional visualization of detectors and physics events. As shown in Fig.~\ref{fig:2D 3D}, an example event of  $\psi(2S)\to\pi^{+}\pi^{-}J/\psi, ~J/\psi\to\gamma\eta_c, ~\eta_c\to\gamma\gamma$ from  Monte Carlo (MC) simulation is displayed. Figure~\ref{fig:2D 3D} (a) shows the two-dimensional display, where the left sud-pad is XY view and the right sub-pad is ZR view. From outer to inner, the pink ring represents the MUC, the blue ring represents the EMC, the yellow ring represents the TOF, and the innermost gray area represents the MDC. The detector units highlighted in red indicate hits fired by the particles, while the black curves and crosses represent reconstructed track information. For example, in Figure~\ref{fig:2D 3D} (a), the two circular tracks in the MDC represent the two oppositely charged pions, and the three hit clusters in EMC without corresponding MDC tracks indicate that they are three photons.

Figure~\ref{fig:2D 3D} (b) is the three-dimensional display. To emphasize visualization of the event, some parts of the sub-detectors are set to invisible. The hits on the two barrel tracks indicate $\pi^{+}\pi^{-}$ interactions with MDC wires. The blue EMC barrel wall shows three red hits that are not $\pi^{+}\pi^{-}$ interactions, corresponding to three photons, where the two large clusters correspond to high-energy decay photons of $\eta_c$, and the small cluster corresponds to low-energy radiated photons of $J/\psi$.
The 2D display is based on a 2D plane projection, while the 3D display is developed using OpenGL in the ROOT framework. This allows users to examine the detectors and events from different perspectives and gain a comprehensive understanding of the spatial distribution and characteristics of the particles and their interactions with the detector.
%BESIII has effectively applied BesVis in data quality monitoring systems, physics simulations, and data reconstruction since operating in 2009. Additionally, due to the intuitiveness of event display software, an increasing number of researchers are willing to use event displays to visually present the physics processes they are studying. For example, they use visualized images in papers, journal covers, and articles to facilitate intuitive analysis. However, the direct application of visualization in specific physics analysis is still limited at present.
%It is necessary to promote the use of visualization in physics analysis.

%%%%%%%%%%%%%%%%%%%
%%%%%%%%%%%%%%%%%%%

Since the development of BesVis, it has played an important role in detector design optimization, detector simulation, reconstruction algorithm tuning, data quality monitoring, the paper publication with journal cover~\cite{10.1093/nsr/nwab201,BESIII:2021cxx}, outreach, and education for particle physics. With the promotion of the visualization method in BESIII, BesVis has also gradually enhanced its application in physics analysis in recent years.

\section{Application in analysis}
\label{sec:application}
In this section, we will take the application of BesVis in the physics analysis of BESIII as an example to introduce the role of visualization in physics analysis, including signal validation, background rejection, and assisting new analysis technique development. The following scenarios will be discussed:
\begin{itemize}
\item Visualization aids in identifying the authenticity of new physics signals in search for invisible decay of $\Lambda$ baryons~\cite{PhysRevD.105.L071101}.
\item Visualization helps in studying background peaks in search for charmonium rare semi-muonic decay $J/\psi\to D^{-}\mu^{+}\nu_{\mu}$~\cite{BESIII:2023fqz}.
\item Visualization assists in further reducing background and improving data sensitivity to new physics in search for lepton flavor violation decay $\psi(nS)\to e^{-}\mu^{+}$~\cite{BESIII:2022exh}. 
\item Visualization demonstrates the potential to implement machine learning techniques for signal discrimination in observation of $\bar{\Lambda}^-_c\to \bar{n} e^{-}\bar{\nu}_{e}$ decays.
\end{itemize}

\subsection{Validating potential discovery in invisible decay of $\Lambda$ baryon}
\label{sec:invisible}

%Current cosmological theories suggest that ordinary matter composed of Standard Model particles only accounts for less than 5\% of the universe, while approximately 26\% is composed of dark matter. 
In collider experiments, it is possible to generate dark matter, which interacts very weakly with ordinary matter, making it invisible in detectors. BESIII operates at the threshold energy point of charm production, its clean background and complete reconstruction of particles enable the search of invisible particles. Searching for the invisible decay of lambda baryons in BESIII is one of the ways to search for dark matter~\cite{PhysRevD.105.L071101}.

\begin{figure*}[tbp]
\centering
\subfigure[]
{\includegraphics[width=0.49\textwidth]{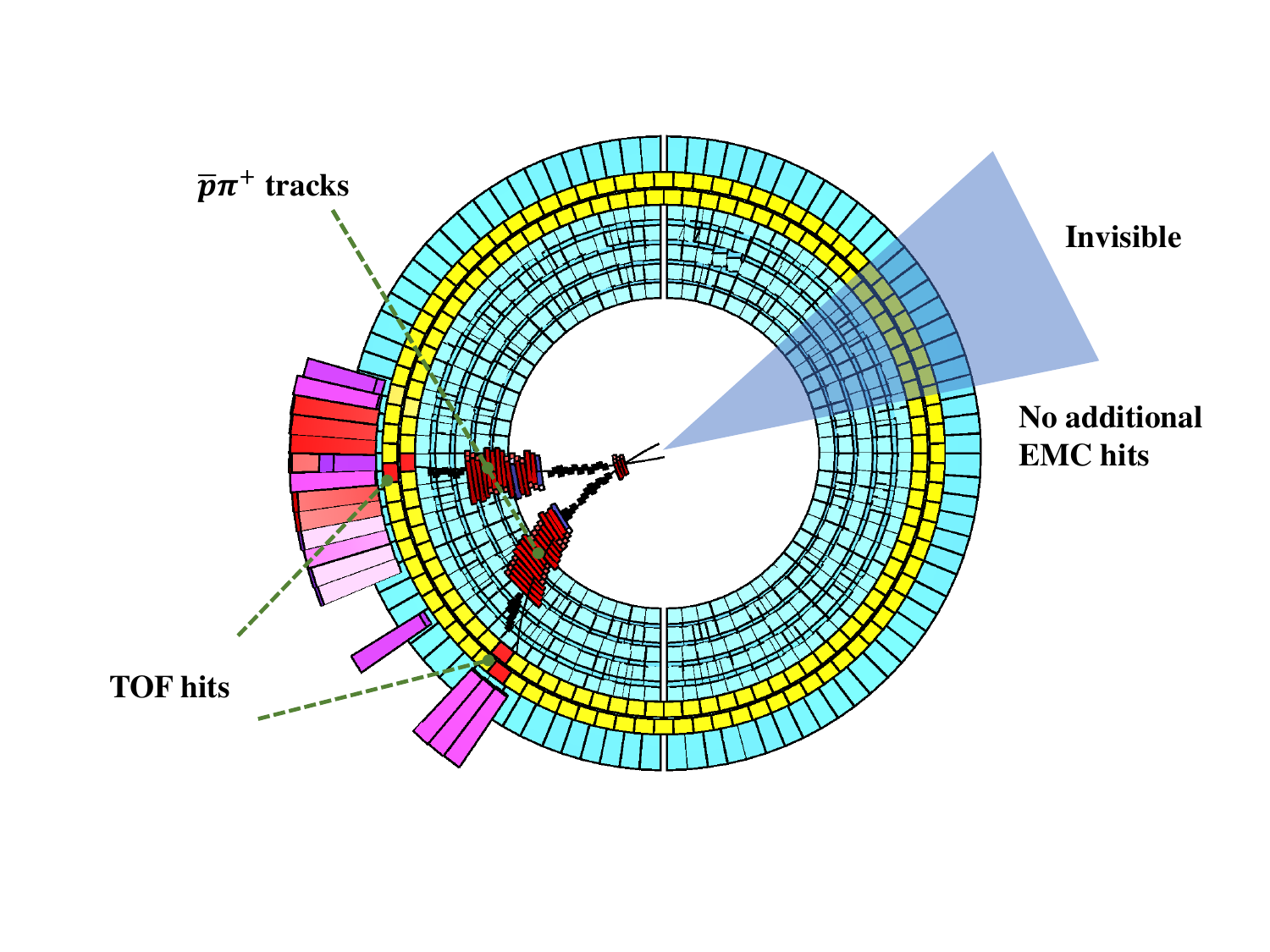}}
\hspace{0.0\textwidth}
\subfigure[]
{\includegraphics[width=0.49\textwidth]{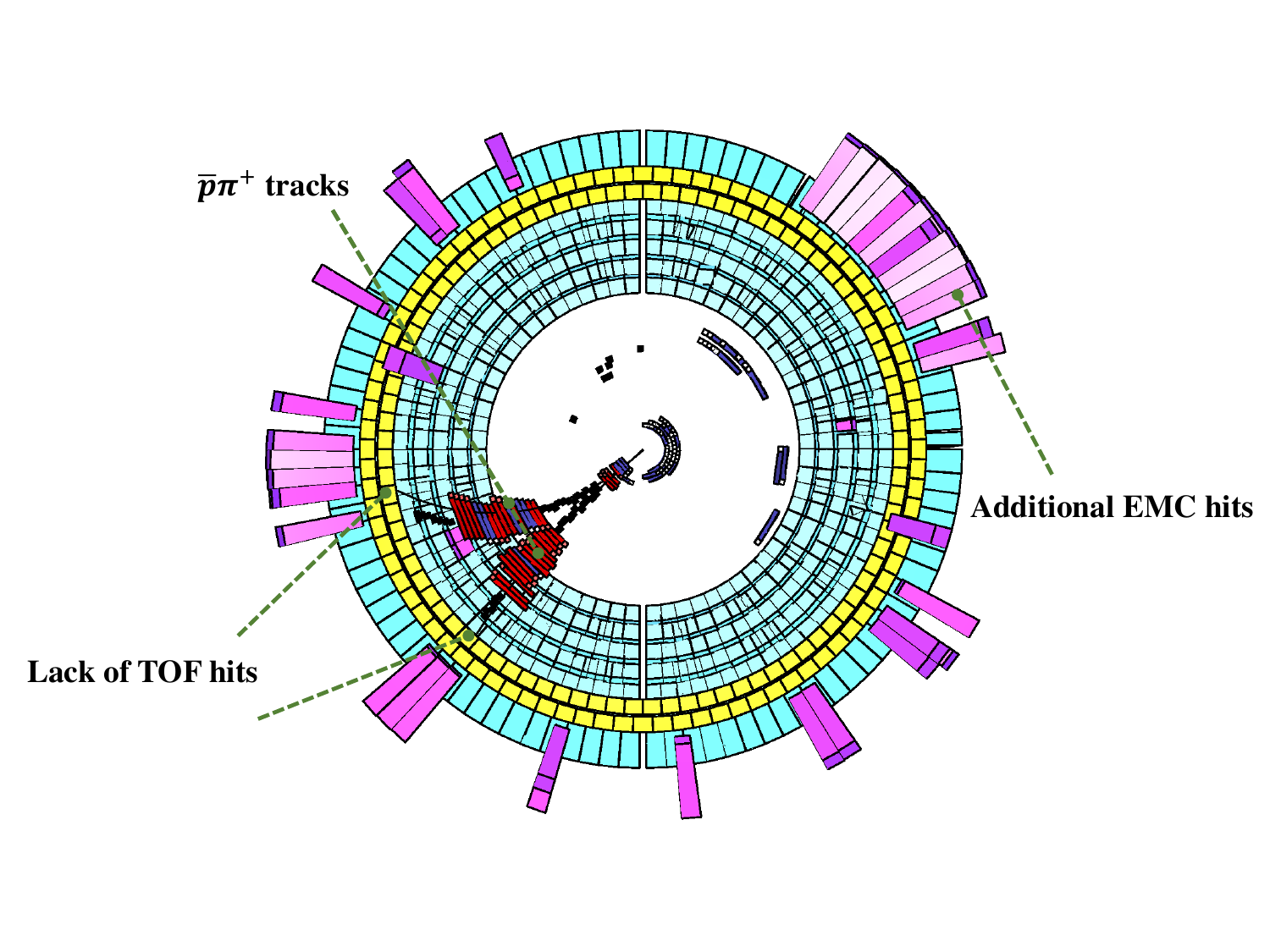}}\\
\caption{(a) An event display simulated for $J/\psi\to\Lambda\bar{\Lambda}, \bar{\Lambda}\to\bar{p}\pi^{+}, \Lambda\to\rm{invisible}$. (b) A fake ``dark matter signal" in observed data from statistical cut-based analysis. The blue region represents EMC, and the yellow region represents TOF. The inner red hits indicate $\bar{p}\pi^{+}$ tracks, while the red squares within the yellow region represent TOF hits. The outermost purple-red bars represent EMC clusters. In (b), the additional EMC hits on the opposite of $\bar{p}\pi^{+}$ indicate that this event is a false signal, and the lack of TOF hits implies the reason for the fake signal.
%In the invisible decay of $\Lambda$ (a), there is no evident hit information on the opposite side of $\bar{p}\pi^{+}$ in the EMC. In "dark matter signal" (b), there is no TOF hits of $\bar{p}\pi^{+}$ and a significant amount of EMC hit information on the opposite side of $\bar{p}\pi^{+}$ is observed.
}
\label{fig:invisible}
\end{figure*}

In BESIII, $\Lambda$ baryons are often produced in pairs. For instance, they can be generated through the decay of $J/\psi$ particles with a 3.097 GeV threshold, resulting in $\Lambda \bar{\Lambda}$ pairs. By tagging $\bar{\Lambda}\to\bar{p}\pi^{+}$, a clean sample of $\Lambda$ on the opposite side can be obtained, allowing researchers to study the invisible decay of $\Lambda$ baryons. 
The energy deposition $E_{EMC}$ of showers in EMC on the opposite side of $\bar{\Lambda}$ can serve as a physical variable to characterize the invisible signal. Since the invisible dark matter does not interact with the detector, the signal of interest will distribute near zero on the $E_{EMC}$ spectrum. Hence, an excess near zero on the $E_{EMC}$ spectrum implies a possible signal from invisible dark matter. After applying traditional statistical cut-based selection criteria, during the early stage of data analysis, a prominent target ``signal'' peak was discovered. 
Using visualization software to display such events beyond expectation with more information can help differentiate the authenticity of the ``dark matter signal''. If the dark matter signal exists, there should be no hits in EMC on the opposite side of $\bar{p}\pi^{+}$, as shown in Figure~\ref{fig:invisible}(a). However, as shown in Figure~~\ref{fig:invisible}(b), a significant amount of hits are observed, which indicates that these events are not genuine dark matter signals. 

%%%%%%%%%%%%%%%%%%%
%%%%%%%%%%%%%%%%%%%

Visualization helps to eliminate the fake signals in the observed data through comprehensive detailed information of the event, and further study in the visualization found that there are no TOF hits of $\bar{p}\pi^{+}$ in the ``dark matter signal'' and the hit time of the additional strange showers in EMC is larger than the expectation, which implies the source of the false signals.
In fact, the previously discovered ``dark matter signal'' was due to a hidden issue with time triggers in the BESIII offline software. In BESIII, photons are usually selected within the time window of 0-700 ns, and the starting time of this window (T0)~\cite{Guan:2013jua}, is usually determined by the triggering of charged particles in MDC and TOF. When no charged particles hit the TOF detector, the uncertainty in T0 can be relatively large, leading to the omission of some photons that should have been detected. This results in the appearance of a fake signal peak near zero in $E_{EMC}$ spectrum.
The ``dark matter signal'' disappeared once the requirement of at least one charged track hitting TOF was imposed~\cite{PhysRevD.105.L071101}. Visualization can help researchers quickly determine whether the events beyond expectation are genuine signals, thus avoiding false discoveries. In comparison to traditional statistical cut-based physical analysis, the visualization approach displays more low-level hit information of events and is less affected by the possible issues during reconstruction and analysis data processing, allowing for direct, intuitive, and quick assessment of the authenticity of observed signals. It holds significant practical value in the search for rare processes and new physics.

\subsection{Studying peaking background in search for semi-muonic charmonium decay}
\label{sec:Dmunv}
The charmonium $J/\psi$ can decay into a charmed meson through weak decay, although its branching ratio is extremely low in the standard model (SM)~\cite{Wang:2007ys,Shen:2008zzb,Dhir:2009rb,Ivanov:2015woa,Wang:2016dkd}. 
Nevertheless, some new physics models beyond the SM can significantly enhance this process by several orders of magnitude~\cite{Datta:1998yq}. Hence, the search for weak decays of $J/\psi$ is a sensitive probe to test SM and explore new physics possibilities~\cite{BESIII:2020nme,Li:2012vk,wang2020new,Chen:2021fcb,BESIII:2022mxl}, such as searching for semi-muonic decay $J/\psi\to D^{-}\mu^{+}\nu_{\mu}$~\cite{BESIII:2023fqz}.

In $J/\psi\to D^{-}\mu^{+}\nu_{\mu}$ analysis, the $D^{-}$ meson is reconstructed using the $D^{-}\to K^{+}\pi^{-}\pi^{-}$ decay channel, so the final state requires identifying four charged tracks corresponding to $K^{+}\pi^{-}\pi^{-} \mu^{+}$, along with an undetectable neutrino carrying missing information.
To characterize the signal of this process, a physical variable $U_{\rm{miss}}$ is defined as $U_{\rm{miss}}=E_{\rm{miss}}-|\vec{P}_{\rm{miss}}|$, where $E_{\rm{miss}}$ represents missing energy, and $|\vec{P}_{\rm{miss}}|$ represents missing momentum, both contributed by the neutrino in the final state. For the signal process $J/\psi\to D^{-}\mu^{+}\nu_{\mu}$, $U_{\rm{miss}}$ follows a Gaussian-like distribution centered around zero. If the semi-muonic weak decay is present in the data, a peak around zero should be observed in the $U_{\rm{miss}}$ spectrum.

%%%%%%%%%%%%%%%%%%%
\begin{figure*}[tbp]
\centering
\subfigure[]
{\includegraphics[width=0.49\textwidth]{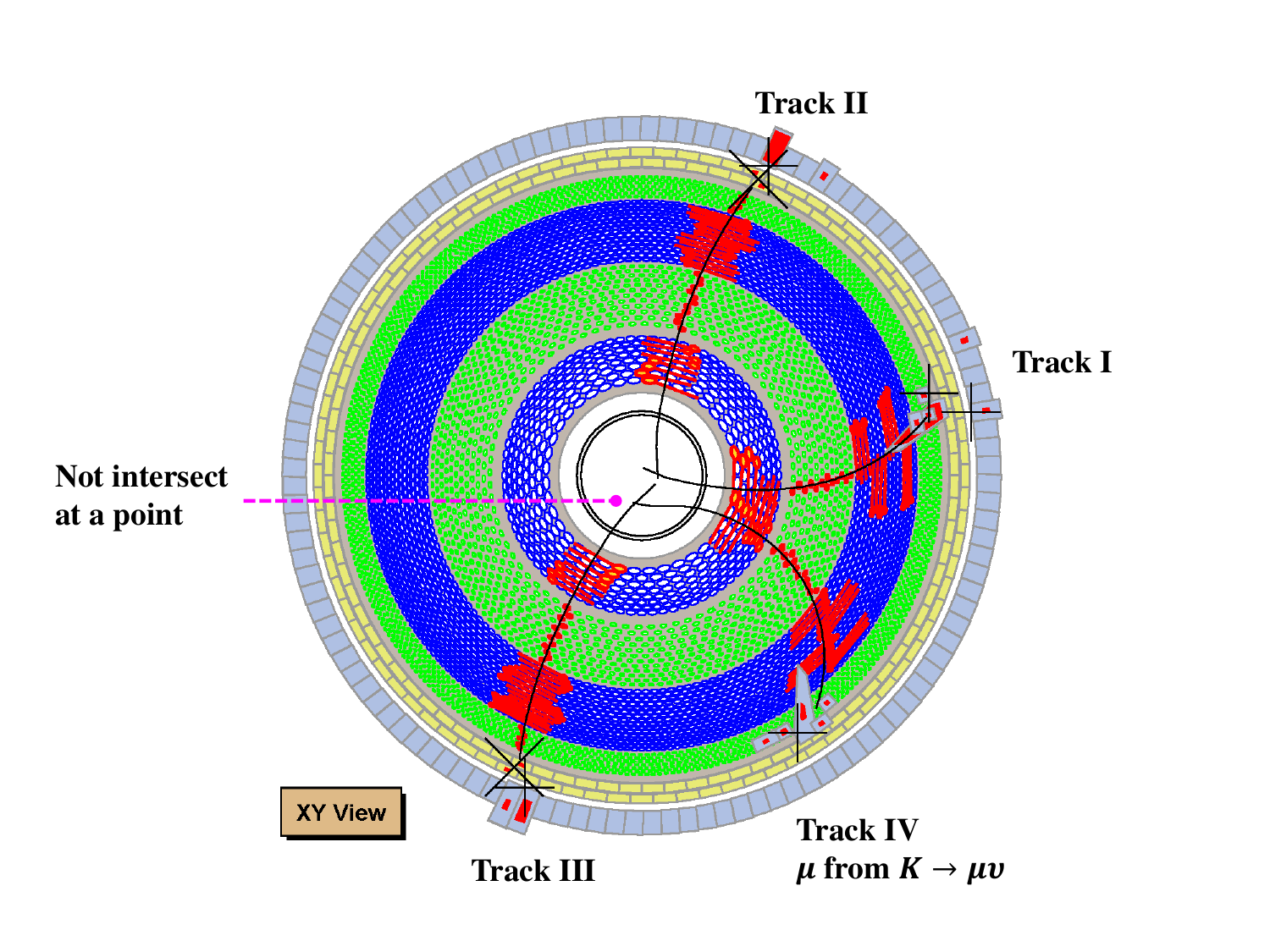}}
\hspace{0.0\textwidth}
\subfigure[]
{\includegraphics[width=0.49\textwidth]{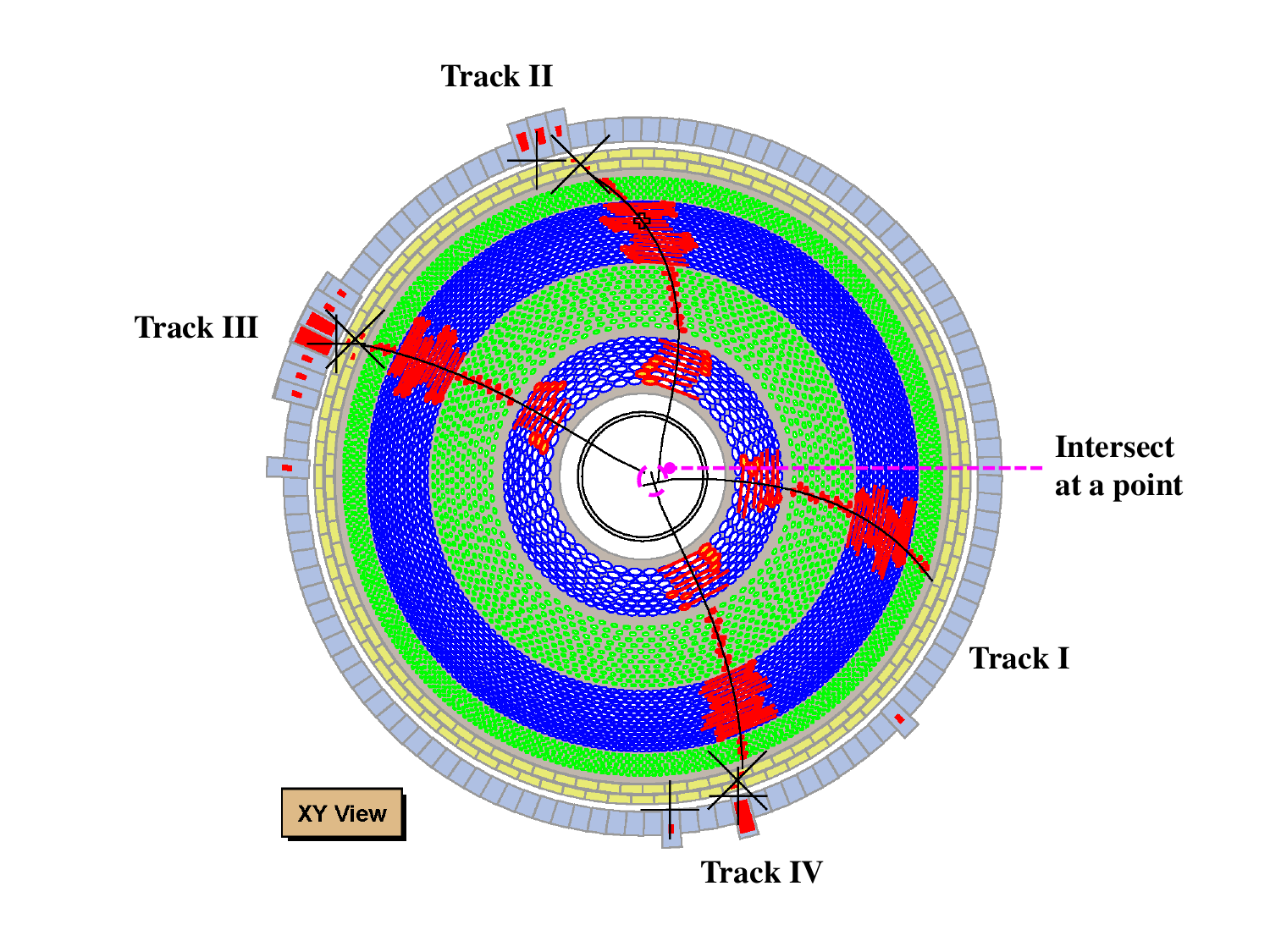}}\\
\caption{(a) $J/\psi\to K^{+}K^{-}\pi^{+}\pi^{-}, K^{-}\to\mu^{-}\bar{\nu}_{\mu}$ background events, where the four charged tracks cannot intersect at a single point. (b) $J/\psi\to D^{-}\mu^{+}\nu_{\mu}, D^{-}\to K^{+}\pi^{-}\pi^{-}$ signal events, where the four charged tracks can intersect at a single point, indicating they are from the same primary vertex. 
%From outer to inner, the light blue ring represents the EMC, the yellow ring represents the TOF, and the innermost green and blue area represents the wires of MDC. The black tracks and crosses represent reconstructed particle information. 
The display mode is called fisheye view, which amplifies the central region for better observation of the particles' vertices.
The green and blue circles represent the stereo and straight wires in MDC, respectively.
}
\label{fig:dmunv}
\end{figure*}
%%%%%%%%%%%%%%%%%%%

Based on the traditional statistical cut-based analysis, a clear peak was found in the $U_{\rm{miss}}$ spectrum near 0. Similarly, visualization can help quickly inspect the physical behavior of this peak event. Using BesVis for event display, it was observed that among the four charged track paths in the final state, only three intersect at the same point near the vertex, while the fourth track deviates, as shown in Figure~\ref{fig:dmunv} (a).
%where the central beam tube has undergone amplification processing, which is called the fisheye mode. 
In contrast, normal $J/\psi\to D^{-}\mu^{+}\nu_{\mu}, D^{-}\to K^{+}\pi^{-}\pi^{-}$ events, where all four charged track are produced near the collision vertex, exhibit all four tracks intersecting at the same point in the visualization, as shown in Figure~\ref{fig:dmunv} (b). 

Visualization of the event shows that one of the charged tracks in the peaking background is not produced in the collision vertex, indicating that the source of the background may come from a secondary decay of one particle in flight.
Actually, these peaking events come from the background $J/\psi\to K^{+}K^{-}\pi^{+}\pi^{-}, K^{-}\to\mu^{-}\bar{\nu}_{\mu}$. 
Although the lifetime of $K^{\pm}$ is relatively long compared to the time scale of an event in BESIII, there are still a very small fraction of $K^{\pm}$ that decay within the selection range of good charged tracks and contribute to the peaking background.
In the event display, the charged track that does not intersect with the other three tracks at a single point is precisely the muon produced from the decay of $K^{\pm}$ after flying a short distance. The visualization method can easily identify the physical characteristics of such peaking background events. Furthermore, it can play a crucial role in suppressing these backgrounds and improving the sensitivity to signals in experimental data ~\cite{BESIII:2023fqz}.

\subsection{Reducing detector associated background in search for charged lepton flavor violating decay}
\label{sec:emu}
$(e,\nu_e)$, $(\mu,\nu_{\mu})$, and $(\tau,\nu_{\tau})$ are three generations of leptons in the SM. Each generation has its own flavor. With the discovery of neutrino oscillation, the flavor violation of neutral leptons has been observed~\cite{Super-Kamiokande:1998kpq,SNO:2002tuh}. However, charged lepton flavor violation (CLFV) is heavily suppressed in the SM. Therefore, the discovery of any CLFV process would be a clear signal of new physics beyond the SM~\cite{Bernstein:2013hba,Cei:2014jtm}. Searching for CLFV process is an important goal in modern particle physics experiments, and one direction is to study it in heavy quark system~\cite{CLEO:2008lxu,BaBar:2021loj,Belle:2022cce,BES:2004jiw,BESIII:2013jau,BESIII:2021slj,BESIII:2022exh}, such as search for $\psi(2S)\to e^\pm\mu^\mp$.

In statistical cut-based physical analysis, it is necessary to select $\psi(2S)\to e^\pm\mu^\mp$ signal events based on the distinct characteristics of electrons and muons. For instance, electrons tend to deposit a majority of their energy in EMC, while muons only deposit a small fraction of their energy in EMC. High-momentum muons with strong penetrating power can reach deep into the outermost layer of MUC, whereas electrons have limited penetrating power and can hardly reach MUC. By exploiting these characteristics, it is possible to effectively suppress the background noise and investigate whether $e\mu$ events are present in the data.
Although the background noise level has been significantly reduced through traditional event selection, further reducing the background is still crucial to enhance the sensitivity for discovering rare new physics processes. The visualization method can be used to further reduce the remaining background.

\begin{figure*}[tbp]
\centering
\subfigure[]
{\includegraphics[width=0.49\textwidth]{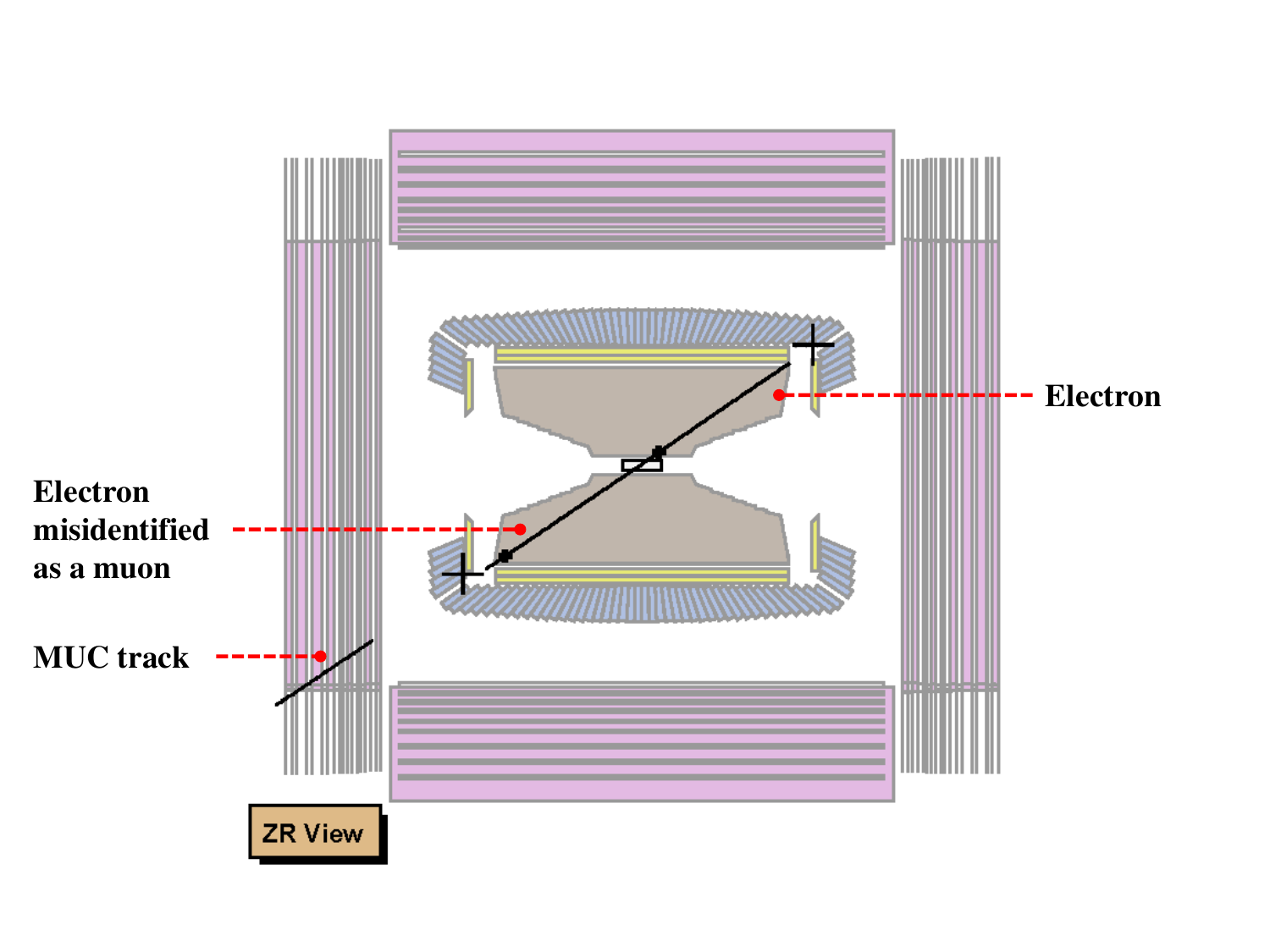}}
\hspace{0.0\textwidth}
\subfigure[]
{\includegraphics[width=0.49\textwidth]{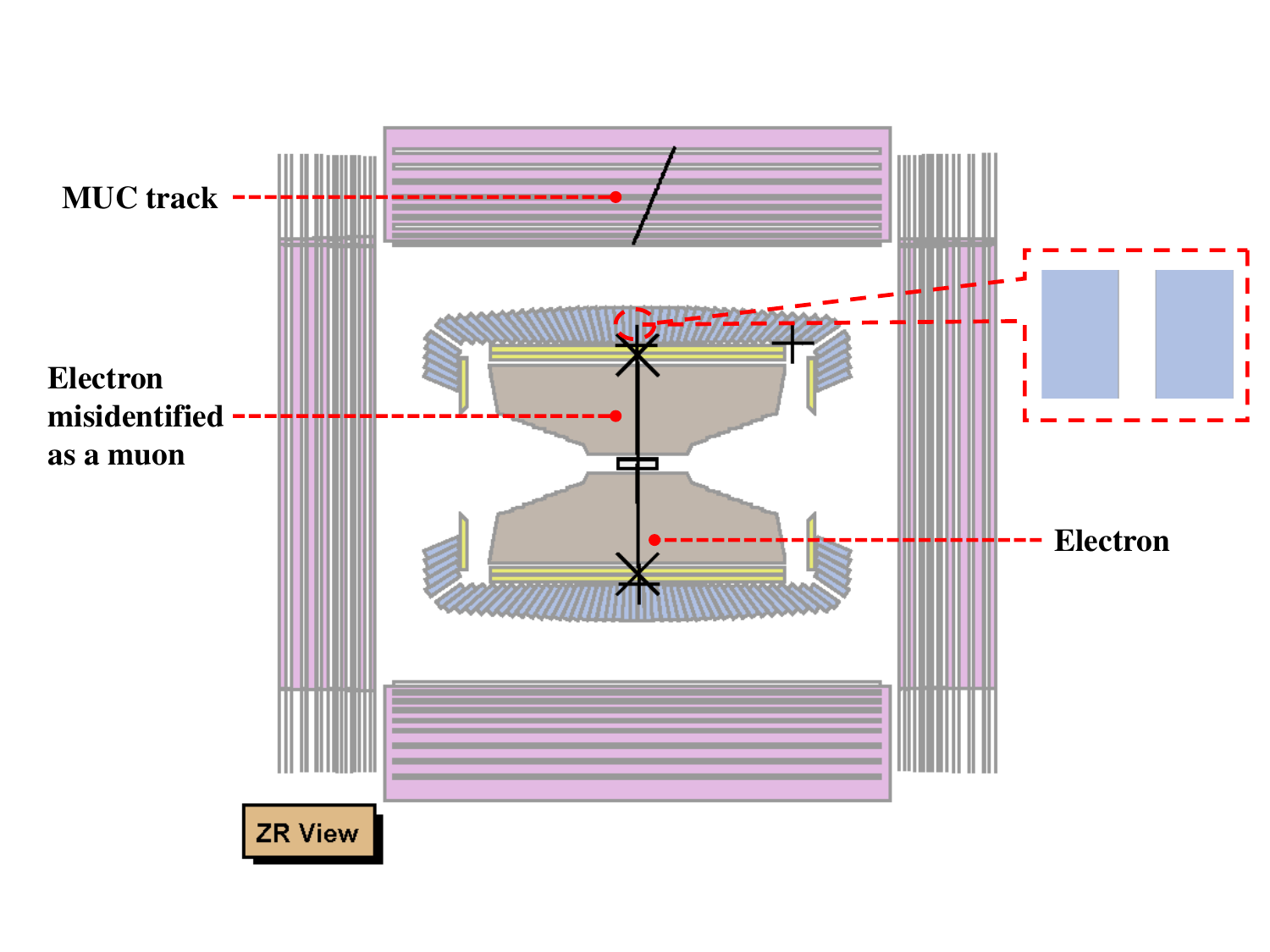}}\\
\caption{Two background events in searching for $\psi(2S)\to e^\pm\mu^\mp$. The black lines represent the reconstructed charged tracks. (a) The particle has a $\rm{cos}\theta\approx0.85$ and passes through the gap between the EMC barrel and endcap. (b) The particle has a $\rm{cos}\theta\approx0$ and passes through a small gap between EMC crystals pointing to the collision vertex.}
\label{fig:emu}
\end{figure*}

After applying visualization to scan the remaining background events (mostly $e^{+}e^{-}\to e^{+}e^{-}$), it is found that these backgrounds are related to the structure of BESIII detector. 
As shown in Figure~\ref{fig:emu}, the direction of the charged track is near $\rm{cos}\theta\approx0.85$ (a) or $\rm{cos}\theta\approx0$ (b). 
Here, $\theta$ represents the polar angle between the direction of the particle and the beam axis.
In these two directions, there are gaps between EMC crystals, the electrons will traverse the gap and deposit only a small fraction of its energy in EMC. Meanwhile, if a cosmic ray muon happens to pass through the MUC and coincide with the electron track in time, or the escaped electron interacts in the outer detector material and produces secondary particles hitting MUC, the electron will be misidentified as a muon track since it satisfies the muon selection criteria, leading to the misidentification of an $e^{+}e^{-}$ background event as an $e^{+}\mu^{-}$ signal event. Although the probability of such coincidence is very low, it becomes a major background while analyzing billions of events, which is a common characteristic in searching for rare processes.

By performing the visualization-based analysis method, the origins of these backgrounds can be discovered, identified, and understood, enabling further suppression of these backgrounds with the polar angle cut. Undoubtedly, it can increase the sensitivity of existing data in search for rare processes, allowing for a better search for new physics or an improved constraint on new physics models.

%%%%%%%%%%%%%%%%%%%
%%%%%%%%%%%%%%%%%%%

\subsection{Studying discrimination of $\bar{n}$ and  $\bar{\Lambda}$ baryon  in $\bar{\Lambda}^-_c$ decay}
\label{sec:nenv}
$\bar{\Lambda}^-_c$ is the lightest charmed baryon and studying its semi-leptonic decays can provide a valuable test for non-perturbative QCD theory~\cite{Li:2021iwf}. The decay width of $\bar{\Lambda}^-_c$ semi-leptonic decays is related to the form factor and the CKM matrix element, where the form factor is a non-perturbative QCD parameter, and CKM matrix element describes quark flavor transition~\cite{Cabibbo:1963yz,Kobayashi:1973fv}. If it is the process $c\to s$, $\bar{\Lambda}^-_c$ can decay into $\bar{\Lambda} l^{-} \bar{\nu}_{l}$ ($l=e,\mu$), while for the process $c\to d$, $\bar{\Lambda}^-_c$ can decay into $\bar{n} l^{-}\bar{\nu}_{l}$. Due to the Cabibbo suppression mechanism, the branching ratio of $\bar{\Lambda}^-_c$ anti-neutron semi-leptonic decay will be much lower than that of the anti-Lambda baryon.

Studying $\bar{\Lambda}^-_c\to \bar{n} l^{-}\bar{\nu}_{l}$ in BESIII presents significant challenges as BESIII lacks a dedicated hadron calorimeter, and detection of neutrons primarily relies on EMC. One key difficulty lies in distinguishing the main background from $\bar{\Lambda}^-_c\to \bar{\Lambda} l^{-} \bar{\nu}_{l}$, where $\bar{\Lambda}\to \bar{n} \pi^0$, $\pi^0\to\gamma\gamma$. The ability of EMC to identify the additional $\pi^0$ from the anti-neutron background will determine the feasibility of this analysis in BESIII. 

\begin{figure*}[tbp]
\centering
\subfigure[]
{\includegraphics[width=0.49\textwidth]{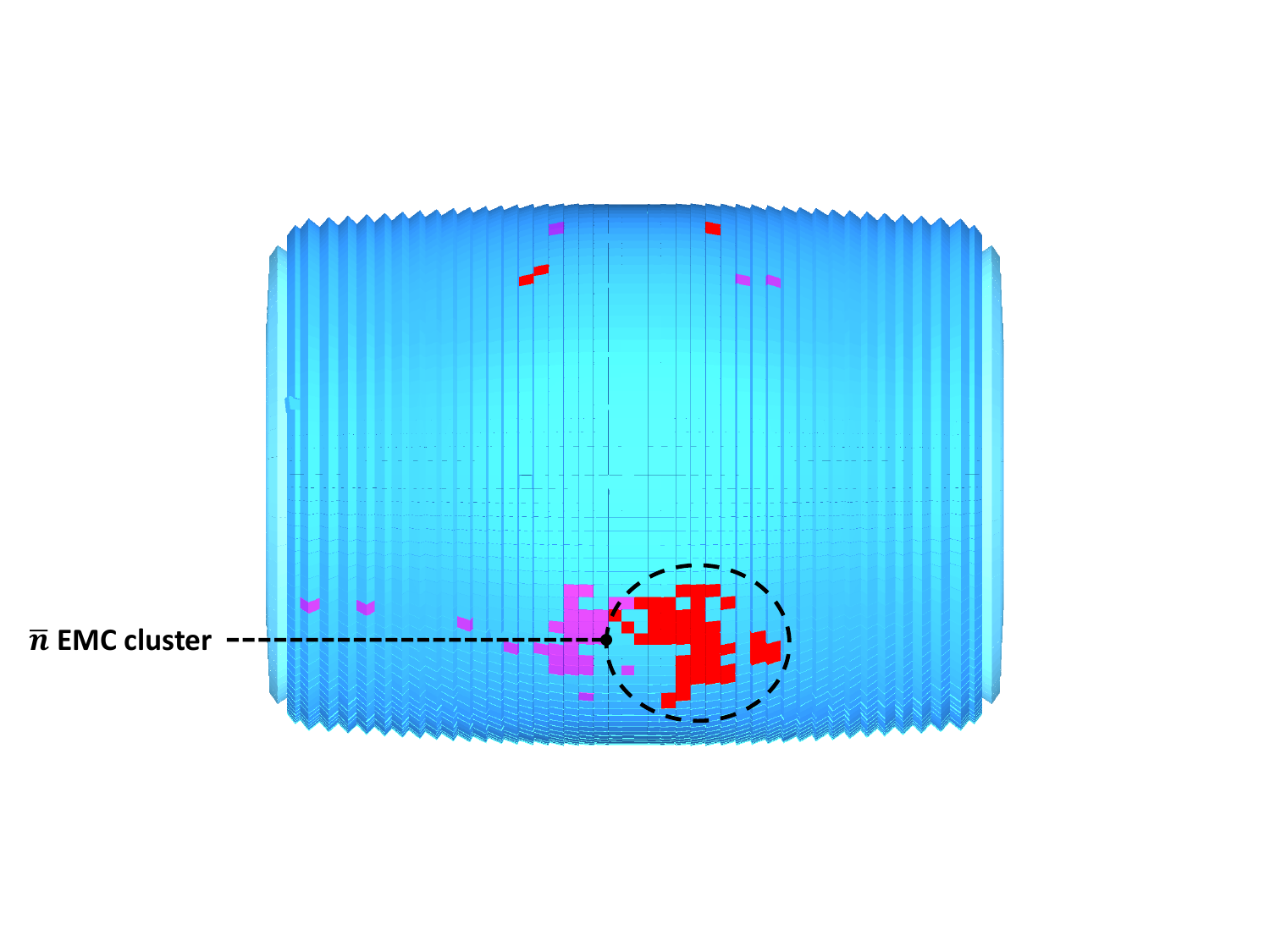}}
\hspace{0.0\textwidth}
\subfigure[]
{\includegraphics[width=0.49\textwidth]{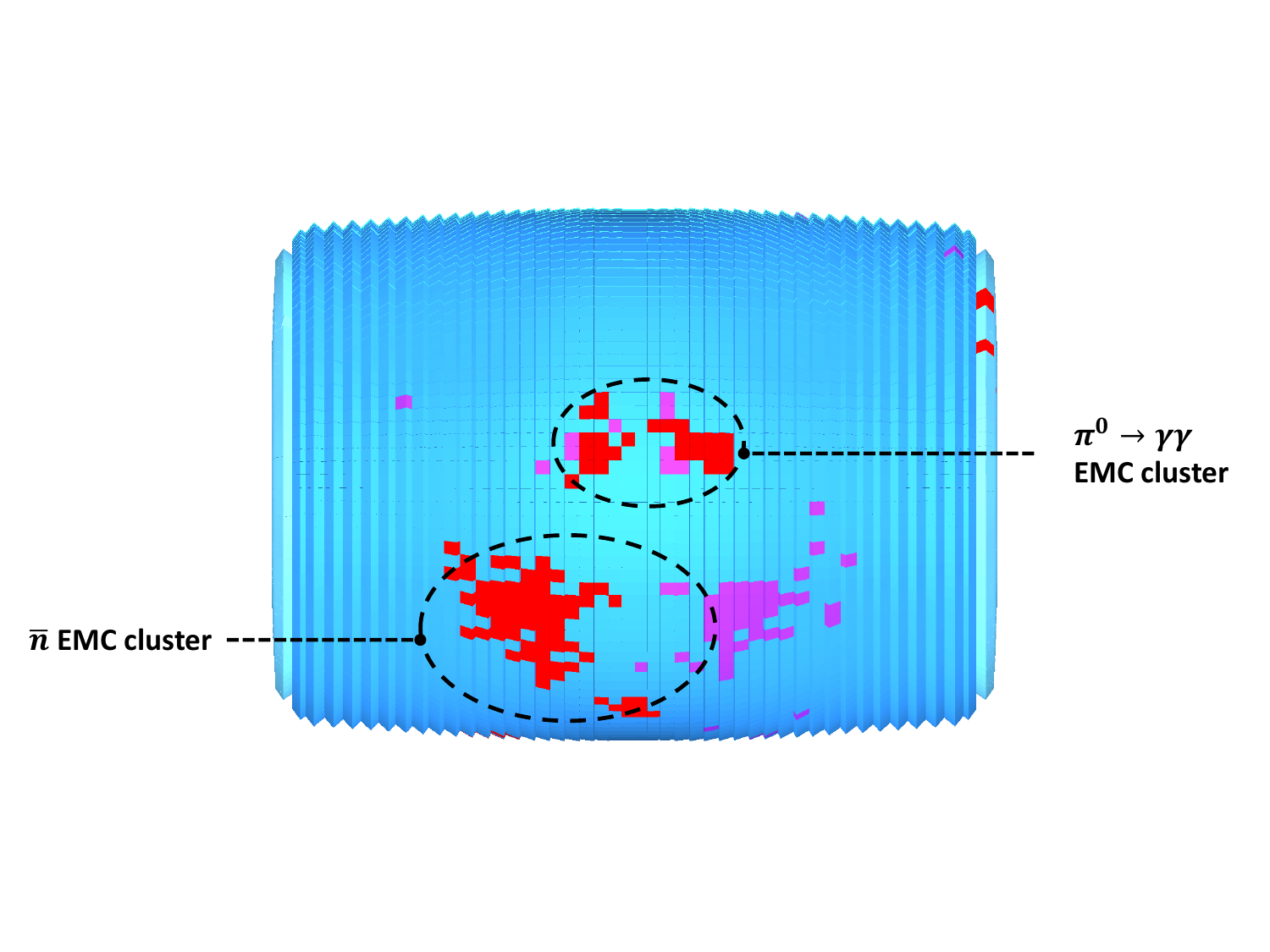}}\\
\caption{Comparison of reaction signals for $\bar{n}$ (a) and $\bar{\Lambda}\to\bar{n}\pi^{0}$ (b) in the EMC. The blue part represents the EMC, and the red parts indicate the hits of neutral particles in the EMC. In (a) and (b), the large red hits in the lower half of the EMC are associated with anti-neutron, while the two smaller red hits in the upper half of the EMC in (b) come from $\pi^{0}\to\gamma\gamma$ decay.}
\label{fig:nenv}
\end{figure*}

Before performing the complex statistical statistical cut-based method in this analysis, the visualization method can be used to intuitively show the differences of EMC cluster shape between anti-neutron and $\bar{\Lambda}\to \bar{n} \pi^0$.
As shown in Figure~\ref{fig:nenv} (a), the reaction of anti-neutron in the EMC often exhibits a larger and more complex hit shape, whereas the two small hits at the top in Figure~\ref{fig:nenv} (b) represent the typical cluster shape of two photons in the $\pi^{0}$ final state. The difference between the two pictures is quite evident, which means distinguishing anti-neutron and anti-Lambda baryon with the EMC cluster shape is feasible. 
However, achieving this task is still highly complex, and practical implementation might require the use of machine learning tools~\cite{Qu:2019gqs,10.1145/3326362,Li:2022tvg,Qian:2021vnh} such as GNN (Graph Neural Network). 
The visualization method can showcase the potential of such advanced techniques before their implementation in complex statistical cut-based analysis.

%%%%%%%%%%%%%%%%%%%
%%%%%%%%%%%%%%%%%%%

\section{Future development}
\label{sec:future}
In the previous section, we demonstrate the advantages of using visualization-based analysis techniques in various physics analyses in BESIII experiment. While there are additional physics applications~\cite{BESIII:2021slj,BESIII:2010tfr,BESIII:2023jem,BESIII:2023rky,BESIII:2021mnd}, they have not been extensively detailed in this paper. In addition to the currently operational particle physics experiments, there are several experiments currently under construction or in the design stage, such as JUNO~\cite{JUNO:2021vlw}, STCF~\cite{Achasov:2023gey}, and CEPC~\cite{CEPCStudyGroup:2018ghi}. It is necessary to develop visualization software tailored for these future experiments, and the visualization method can also be applied to the upcoming physics analysis work. For example, if the search for CLFV process like $e^+e^-\to e^-\mu^+$ is going to be conducted at STCF or CEPC, the background mentioned in Sec.~\ref{sec:emu} may potentially become a spurious fake new physics signal. In that case, it will be necessary to employ careful event display to validate the authenticity of the new physics discovery.
JUNO is a large-scale neutrino experiment aiming at the determination of neutrino mass hierarchy, in which the background rate is more than six orders of magnitude higher than the reactor neutrino signal rate. The visualization tools~\cite{You:2017zfr,Zhu:2018mzu,Zhang:2020jkg} have been developed in JUNO and are expected to play an essential role in screening the rare signal events for neutrino analyses.

In future developments, it is important that visualization software can keep up with the advancements of software and computing technology from industry to serve physics analysis better. Having a unified platform that can provide consistent visualization services for different experiments would be a favorable choice, including a universal detector description method~\cite{frank_markus_2018_1464634}, a comprehensive detector geometry transformation interface~\cite{Huang:2022wuo}, and a common event data model~\cite{EDM4hep}. %However, different experiments have their own unique geometry files, and traditional geometry files cannot be directly implemented in Unity. To achieve this, an interface needs to be created that can convert the geometry files from various major particle physics experiments into a format compatible with Unity's visualization capabilities~\cite{Huang:2022wuo}. 
Furthermore, if visualization could be utilized through web browsers, its applications would become even more convenient. With such an implementation, visualization will become more versatile and accessible, thus offering better assistance for particle physics analysis.

\section{Summarry}
\label{sec:sum}
Visualization can play a unique role in modern particle physics data analysis. In this manuscript, we introduce the potential value of visualization method for improvement of physics analysis.
The applications of event display in several BESIII physics analyses have demonstrated its versatility and complementarity to the traditional statistical analysis method.
%In the search for the invisible decay of $\Lambda$ baryons, visualization aids in identifying ``new physics signals'' and prevents false reporting of the new physics. In the search for charmonium semi-muonic decays, visualization analyzes the source of peaking backgrounds, facilitating their removal from the analysis. In the search for charged lepton flavor-violating processes, visualization reveals additional features of residual backgrounds, providing a cleaner background environment for the analysis. In the measurement of $\bar{\Lambda}^-_c$ baryon semi-leptonic decay with an anti-neutron, visualization demonstrates the distinction of the signals between anti-neutron and anti-Lambda baryon background, offering a possibility for this measurement. 
%In summary, the visualization method can be combined with the statistical cut-based method, having great potential in physics analysis improvement. 
It is recommended that the visualization method be generally taken in physics analysis, especially in search for rare physics signals in the current operational experiments and design of the next generation particle physics experiments.

\section*{Acknowledgements}

This work was supported by the National Natural Science Foundation of China (Grant Nos. 12175321, 12150005, 11975021, 11675275, and U1932101), National Key Research and Development Program of China (Nos. 2023YFA1606000, 2020YFA0406300 and 2020YFA0406400), State Key Laboratory of Nuclear Physics and Technology, Peking University  (Nos. NPT2020KFY04 and NPT2020KFY05), Chinese Academy of Sciences (CAS) Large-Scale Scientific Facility Program; National College Students Innovation and Entrepreneurship Training Program, Sun Yat-sen University.

%\end{linenumbers}


%\clearpage
\bibliography{mybib}% common bib file

\end{document}